\documentclass[aps,pra,superscriptaddress,reprint,floatfix,longbibliography,showpacs]{revtex4-1}
\usepackage[english]{babel}
\usepackage{amsmath}
\usepackage{amssymb}
\usepackage{graphicx}
\usepackage{upgreek}
\usepackage{epstopdf}
\usepackage{color}
\usepackage{dcolumn}
\usepackage{multirow}
\usepackage[note-name=, use-sort-key = false]{notes2bib}
\usepackage{ulem}

\usepackage[colorlinks,bookmarks=false,citecolor=darkblue,linkcolor=red,urlcolor=blue]{hyperref}
\definecolor{darkred}{rgb}{0.7,0.0,0.0}
\definecolor{darkblue}{rgb}{0,0.02,0.45}

\def\cm{cm$^{-1}$}

\begin{document}
\title{Optical signatures of type-II Weyl fermions in the noncentrosymmetric semimetals \\ $R$AlSi ($R$=La, Ce, Pr, Nd, Sm)}

\author{J. Kunze}
\affiliation{Experimental Physics II, Institute of Physics, University of Augsburg, 86159 Augsburg, Germany}
\author{M. K\"opf}
\affiliation{Experimental Physics II, Institute of Physics, University of Augsburg, 86159 Augsburg, Germany}

\author{Weizheng Cao}
\email{caowzh@shanghaitech.edu.cn}
\affiliation{School of Physical Science and Technology, ShanghaiTech University, Shanghai 201210, China}
\author{Yanpeng Qi}
\email{qiyp@shanghaitech.edu.cn}
\affiliation{School of Physical Science and Technology, ShanghaiTech University, Shanghai 201210, China}
\affiliation{ShanghaiTech Laboratory for Topological Physics, ShanghaiTech University, Shanghai 201210, China}
\affiliation{Shanghai Key Laboratory of High-resolution Electron Microscopy and ShanghaiTech Laboratory for Topological Physics, ShanghaiTech University, Shanghai 201210, China}
\author{C. A. Kuntscher}
\email{christine.kuntscher@physik.uni-augsburg.de}
\affiliation{Experimental Physics II, Institute of Physics, University of Augsburg, 86159 Augsburg, Germany}

\begin{abstract}
Weyl semimetals with magnetic ordering provide a promising platform for the investigation of rare topological effects such as the anomalous Hall effect, resulting from the interplay of nontrivial bands with various spin configurations. The materials $R$AlSi, where $R$ represents a rare-earth element, are prominent representatives of Weyl semimetals, where the Weyl states are
induced by space inversion symmetry breaking and in addition, for several rare-earth elements $R$, enhanced by time-reversal symmetry breaking through the formation of a magnetic order at low temperature. We report optical signatures of Weyl fermions in the magnetic compounds CeAlSi, PrAlSi, NdAlSi, and SmAlSi as well as the non-magnetic family member LaAlSi by broad-frequency infrared spectroscopy at room temperature, i.e., in the paramagnetic phase. A similar profile of the optical conductivity spectrum and a metallic character are observed for all compounds, with LaAlSi showing the strongest free charge carrier contribution. Furthermore, the linear-in-frequency behavior of the optical conductivity of all investigated compounds indicates the presence of Weyl nodes in close vicinity of the Fermi energy, resulting from inversion symmetry breaking in noncentrosymmetric structures. According to the characteristics of these linear slopes, the $R$AlSi compounds are expected to host mainly type-II Weyl states with overtilted Weyl cones. The results are compared to the optical response of the closely related $R$AlGe materials, which are considered as potential hybridization-driven Weyl-Kondo systems.
\end{abstract}

\maketitle

\section{Introduction}
In recent years, the investigation of topological phases in condensed matter has attracted a lot of interest. A prominent example is the Dirac semimetal phase, where linear band crossings occur near the Fermi energy $E_F$, which are protected by spatial and/or time-reversal symmetry~\cite{Armitage.2018}.
Other than high-symmetric Dirac points, the existence of Weyl points requires either inversion or time-reversal symmetry to be broken, creating pairs of Weyl points with different chirality~\cite{Yan.2017}. One famous representative of such Weyl semimetals (WSMs) with broken inversion symmetry is the noncentrosymmetric and nonmagnetic compound TaAs~\cite{Lv.2015}.
When the degeneracy of the Dirac cones is lifted by the broken inversion symmetry, the Weyl cones are shifted in energy relatively to each other, whereas breaking of time-reversal symmetry through magnetic moments, causes a shift in $k$-space~\cite{Soh.2019, Tabert.2016a}. WSMs with broken time-reversal symmetry are rare as compared to those with broken inversion symmetry, yet promising since in inversion symmetric systems there is no shift of the nodes in energy, resulting in a vanishing density of states at the Fermi level~\cite{Soh.2019}. In addition, the combination of topological phases and magnetism enables the realization of rare quantum mechanical effects, such as the intrinsic anomalous Hall effect or the topological Hall effect~\cite{Yang.2021, Wang.2022}. EuCd$_2$As$_2$ represents this group as a famous member, where the alignment of Eu spins magnetically induces Weyl nodes via exchange coupling to create a magnetic WSM candidate~\cite{Soh.2019}.

The family of rare-earth based compounds $R$Al$X$ ($R$ = La, Ce, Pr, Nd, Sm; $X$=Si and Ge) also has attracted a lot of attention recently as WSM candidates~\cite{Lyu.2020, Cheng.2023, Wu.2023, Su.2021,Xu.2017,Dhital.2023}. The noncentrosymmetric $R$Al$X$ family allows the investigation of WSM states in magnetic and nonmagnetic phases in isostructual compounds, as the magnetic contributions depend on the rare-earth element $R$~\cite{Su.2021}. Among the La, Ce, Pr, Nd and Sm elements, only La does not exhibit a magnetic moment in the $R$AlSi compounds~\cite{Wang.2021}. Therefore, varying the rare-earth ion affects the electronic topology through different magnetic ground states~\cite{Cao.2022}. In the case of PrAlSi, a ferromagnetic order sets in at $T_{\mathrm{C}}=17.8\,$K, followed by two spin-glass-like reentrant magnetic transitions at lower temperatures~\cite{Lyu.2020}. There are also several critical temperatures for NdAlSi, as this compound undergoes an antiferromagnetic transition at $T_{\mathrm{N}}=7.2\,$K and a ferrimagnetic transition at $T_{\mathrm{C}}=3.3\,$K~\cite{Gaudet.2021, Wang.2022}.  SmAlSi orders antiferromagnetically at $T_{\mathrm{N}}=10.7\,$K, while CeAlSi also shows a ferromagntic magnetic structure below $T_{\mathrm{C}}=17.8\,$K~\cite{Cao.2022}. $R$AlSi compounds crystallize for the most part in the space group $I$4$_1md$, yet with a certain percentage also the $I$4$_1/amd$ phase has been found experimentally~\cite{Wang.2021}. In the main space group $I$4$_1md$, the unit cell parameters of $R$AlSi vary between $a=4.1566\,$\AA\ for Sm and $a=4.3069\,$\AA\ for La, and $c=14.4552\,$\AA\ for Sm and $c=14.6494\,$\AA\ for La~\cite{Wang.2021}.

Regarding the electronic structure, the $R$Al$X$ compounds have been shown to be promising candidates for the WSM phase, as numerous Weyl points are reported to exist near the Fermi level, e.g., 40 Weyl points
in the paramagnetic phase of NdAlSi~\cite{Wang.2022}, CeAlSi \cite{Sakhya.2023}, and LaAlGe \cite{Xu.2017} upon inclusion of spin-orbit coupling.
The electronic band structure of $R$AlSi is very similar among the different compounds and the density of states at the Fermi level is expected to be vanishingly small, which leads to a semi-metallic ground state~\cite{Lou.2023, Su.2021, Yang.2021, Gaudet.2021}.
Resistivity measurements have confirmed the (semi-)metallic behaviour of the $R$AlSi materials as dc conductivity values between 11000 and 18000 $\Omega^{-1}\mathrm{cm}^{-1}$ have been found in recent publications~\cite{Lyu.2020, Cao.2022, Wang.2022}.

The materials CeAlGe and PrAlGe are furthermore discussed as potential hybridization-driven Weyl-Kondo semimetals \cite{Corasaniti.2021,Yang.2022}. According to optical conductivity studies, electronic correlations due to the hybridization of the localized f electrons of the rare-earth element (Ce 4f$^1$ and Pr 4f$^2$) with the conduction electron states lead to a reduction of the Fermi velocity and an enhancement of the charge carriers effective mass, in contrast to LaAlGe lacking f electrons (La 4f$^0$). It was furthermore suggested that the onset of magnetic ordering in CeAlGe and PrAlGe significantly affects the electronic band structure, as observed in the temperature-dependent optical response \cite{Corasaniti.2021,Yang.2022}, suggesting a coupling between magnetism and the electronic degrees of freedom. In contrast, recent angle-resolved photoemission spectroscopy results combined with first-principles calculations for PrAlSi and SmAlSi revealed only a weak coupling between the localized 4f electrons and the conduction electrons and a negligible effect of the magnetic ordering on the electronic band structure \cite{Lou.2023}. Accordingly, whether the $R$Al$X$ are indeed Weyl-Kondo systems is still a matter of debate.

In this work, the optical response of $R$AlSi ($R$ = La, Ce, Pr, Nd, Sm) is investigated by comparing the optical functions such as the optical conductivity and dielectric function of the various rare-earth based compounds. The results are discussed in terms of the transport and topological properties discussed in the literature~\cite{Lyu.2020, Cao.2022, Wang.2022, Tabert.2016a}. In particular, a comparison to the optical conductivity of the closely related $R$AlGe materials is given.

\section{Methods}

Single crystals of $R$AlSi were synthesized using the self-flux method, as described elsewhere~\cite{Cao.2022a,Cao.2022}.
The dimensions of these plate-like samples are in the range of 0.5 and 3\,mm and high-quality surfaces have been prepared by careful polishing, as they are essential for quantitative infrared reflectivity measurements. We have performed these reflectivity measurements from 100 to 20000\,\cm\ (0.01 to 2.48\,eV) at room temperature using Fourier-Transform Infrared spectroscopy (FTIR) with the use of a Bruker 80v spectrometer, which is coupled to a Hyperion microscope. In order to obtain the absolute reflectance of our samples, these have been aligned perpendicular to the incoming IR beam, which also holds for an aluminum mirror next to it serving as reference. In the case of LaAlSi, additional low-temperature reflectivity measurements were carried out.
By using resistivity values and volumetric data from the literature~\cite{Lyu.2020, Cao.2022, Wang.2022}, these reflectance data were extrapolated in the low and high energy ranges, which is necessary to calculate the optical functions, such as the optical conductivity $\sigma_1$, the real part of the dielectric function $\varepsilon_1$, and the loss function $-\mathrm{Im}(1/\varepsilon)$, from Kramers-Kronig relations using programs by David Tanner~\cite{Tanner.2015}.
The optical functions have been analyzed with the RefFIT software~\cite{Kuzmenko.2005} by fitting with the Drude-Lorentz model.

\section{Results and Discussion}

\begin{figure}[t]
	\includegraphics[width=1\linewidth]{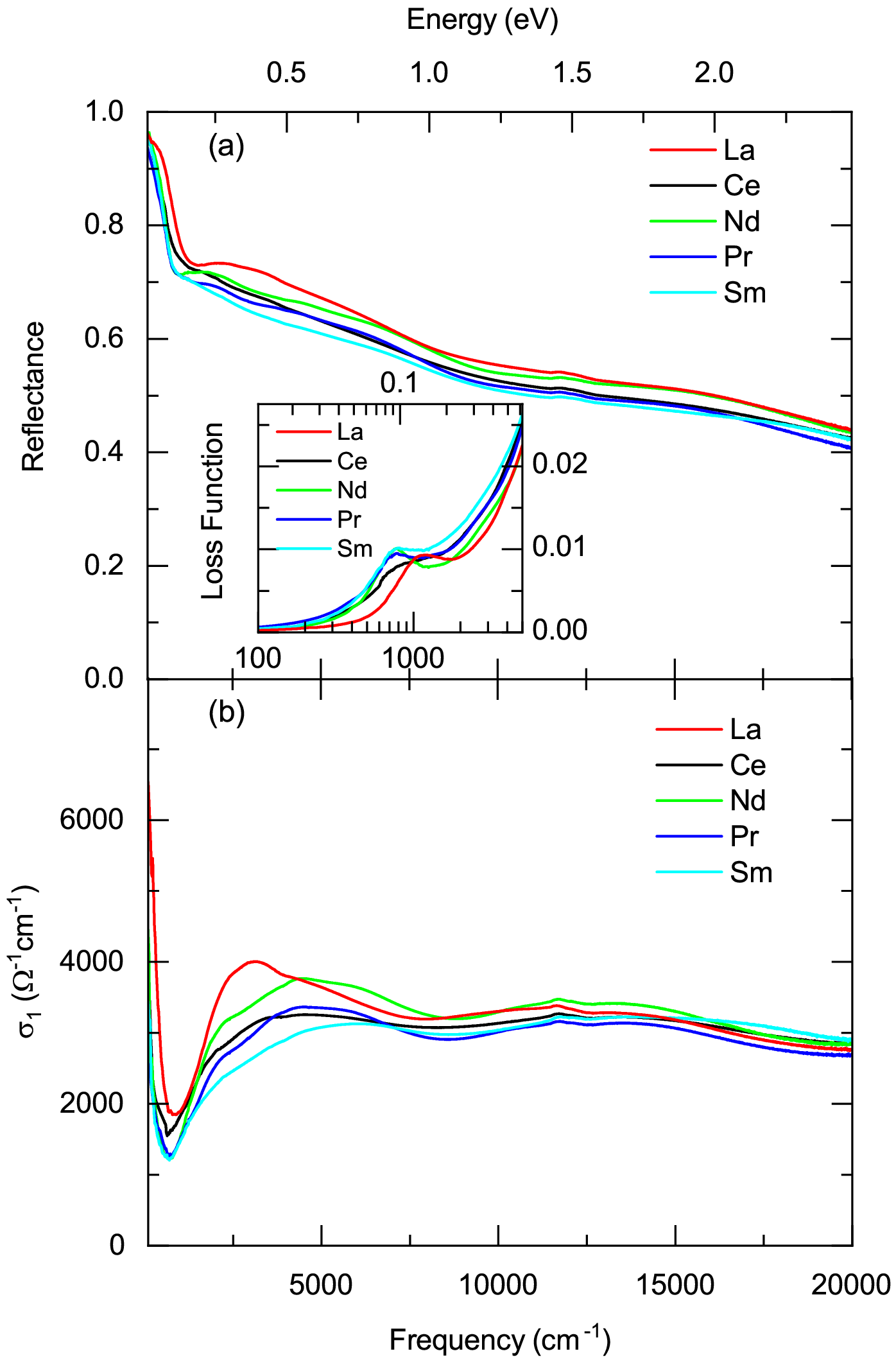}
	\caption{(a) Room-temperature reflectance of five different compounds of the $R$AlSi family ($R$ = La, Ce, Nd, Pr, Sm) at room temperature. Inset: Loss function from 100 to 5000\,\cm\ as an inset. (b) Corresponding optical conductivity $\sigma_1$ for all measured compounds.}
	\label{fig.reflectivity}
\end{figure}

The reflectance spectra displayed in Fig.\ \ref{fig.reflectivity}(a) show a high value towards the low-energy range, indicating the metallic characteristics of the investigated materials of the $R$AlSi compound family. Also, a plasma edge is observed at around 900\,\cm\ for all five compounds. For LaAlSi the plasma edge is located at the highest frequency, indicating that it has the strongest metallic character among the studied samples (see also the discussion of the plasmon peak position below). Towards higher energy, the reflectance spectra show a very similar behavior for all compounds and are just varying slightly in their absolute value, therefore hinting at a similar electronic structure of the $R$AlSi materials in the higher energy regime. Especially noticeable is the merging of the reflectance spectra from approximately 16000 to 20000\,\cm.
Only the compound LaAlSi shows a broad bump at around 2500\,\cm, and hence marks an exceptional behavior among the studied $R$AlSi materials.

\begin{figure}[t]
	\includegraphics[width=1\linewidth]{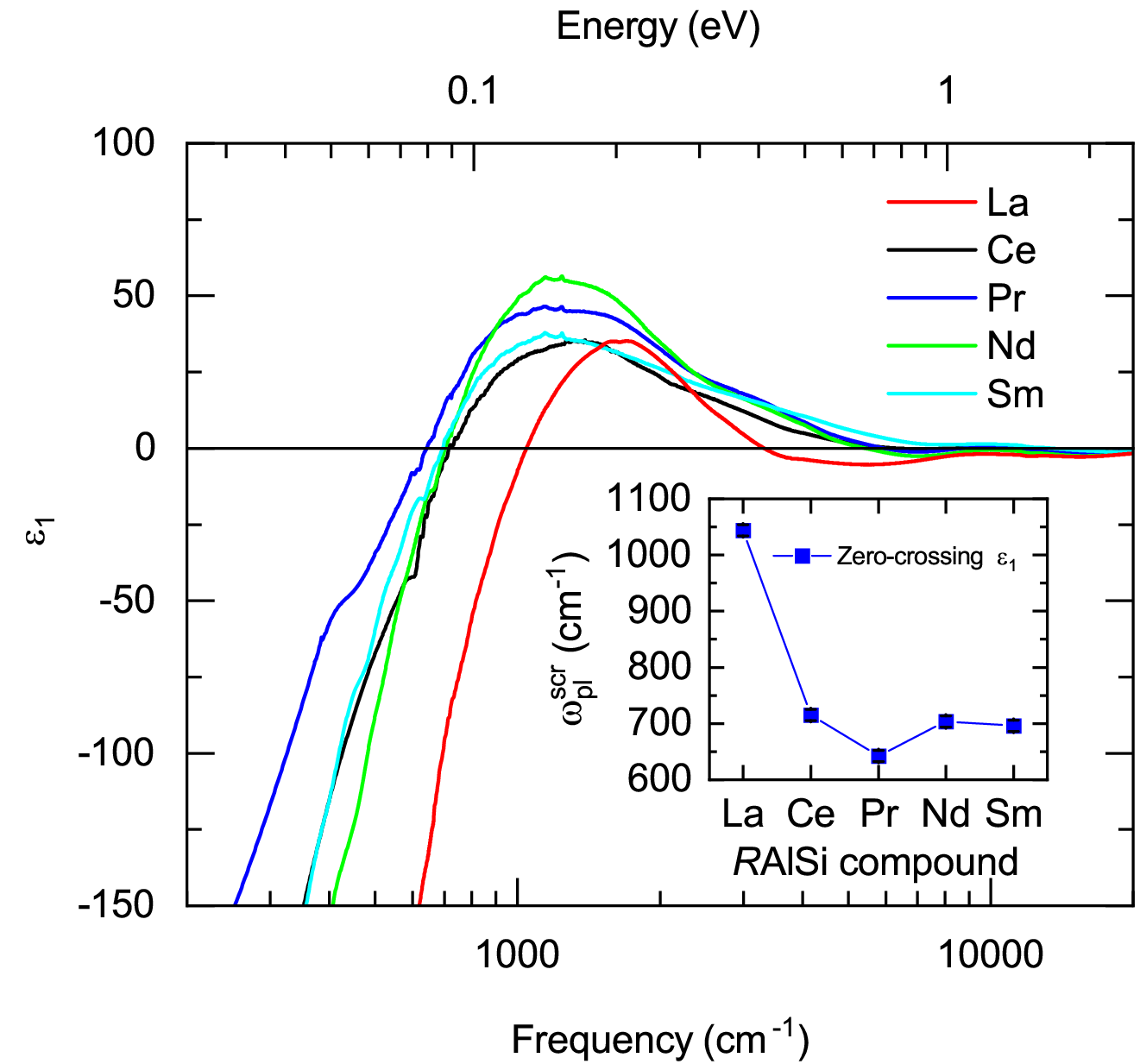}
	\caption{Real part of the dielectric function $\varepsilon_1$ for all studied $R$AlSi compounds at room temperature. The inset shows the values of the screened plasma frequency $\omega_{\mathrm{pl}}^{scr}$ determined by the zero-crossing of $\varepsilon_1$.}
	\label{fig.epsilon1}
\end{figure}

The corresponding optical conductivity spectra $\sigma_1$ are depicted in Fig.\ \ref{fig.reflectivity}(b).
It is important to note that the optical conductivity spectra have a similar profile for all five compounds with only quantitative differences. This is in contrast to results reported for the closely related materials LaAlGe, CeAlGe, and PrAlGe, which exhibit different optical conductivity profiles already at room temperature \cite{Corasaniti.2021,Yang.2022}.
Among the studied compounds, LaAlSi has the highest $\sigma_1$ value at the lowest studied wavenumber 100\,\cm. NdAlSi exhibits the second highest value and CeAlSi and SmAlSi do have very similar values at the low-frequency limit, although the value of CeAlSi is slightly higher, and PrAlSi has the lowest value. Besides the difference for the value of $\sigma_1$ at low energy (100\,\cm\ or 0.01\,eV) and the varying distinction of the first peak in the spectrum (2000 to 5500\,\cm), the spectra of the five different compounds resemble a similar profile towards higher energy. This leads to the conclusion, that the main differences in the materials lie in the low-energy regime, where the excitations of free charge carriers (Drude contribution) and the interband transitions between electronic bands in direct vicinity of the Fermi energy contribute.
The Lorentz contributions in the higher-energy range are very similar for all investigated materials and show no dependence on the rare earth element $R$.

As the reflectance and optical conductivity data already suggest, the compound LaAlSi stands out in comparison to the other four studied compounds.
This is also revealed by the position of the plasmon peak in the loss function (LF) [see inset in Figure~\ref{fig.reflectivity}(a)]. Although no absolute value can be obtained for the position of the plasmon peak, since the peak is not distinct enough, for the compounds of CeAlSi, NdAlSi, PrAlSi and SmAlSi the plasmon peak is located at approximately the same frequency. However, for LaAlSi the peak position is obviously shifted to a higher value. The presence of a plasmon peak in the loss function also indicates the metallic character of all studied compounds.

In general, the position of the plasmon peak yields a value for the screened plasma frequency $\omega_{\mathrm{pl}}^{\mathrm{scr}}$, which can serve as a measure for the metallic strength of a material.
The value of $\omega_{\mathrm{pl}}^{\mathrm{scr}}$ can also be extracted from the zero crossing of the real part of the dielectric function $\varepsilon_1$ (see Fig.\ \ref{fig.epsilon1}).
The so-obtained $\omega_{\mathrm{pl}}^{\mathrm{scr}}$ values for all studied materials are depicted in the inset of Fig.\ \ref{fig.epsilon1}. Firstly, it can be seen that for low frequencies ( $<$ 1000\,\cm) $\varepsilon_1$ takes large negative values, evidencing the metallic character of the investigated materials. Secondly, the negative values of $\varepsilon_1$ at high energies are caused by the rather pronounced interband transitions due to the presence of multiple energy bands in the vicinity of E$_F$. Thirdly,
it is obvious that the screened plasma frequency is the highest for LaAlSi, since the zero crossing point is clearly set apart from the other four compounds towards higher energies. For the compounds CeAlSi, PrAlSi, NdAlSi, and SmAlSi the frequency of the zero crossing is roughly at the same position.
The $\omega_{\mathrm{pl}}^{\mathrm{scr}}$ value for LaAlSi is approximately 1500\,\cm, whereas for the compounds CeAlSi, NdAlSi and SmAlSi, a value of around 700 \,\cm\ is obtained. The compound PrAlSi has the smalles value of 650 \,\cm. Most importantly, LaAlSi has by far the highest value of the screened plasma frequency, which is in accordance with the loss function, reflectance, and optical conductivity.

\begin{figure}[t]
	\includegraphics[width=1\linewidth]{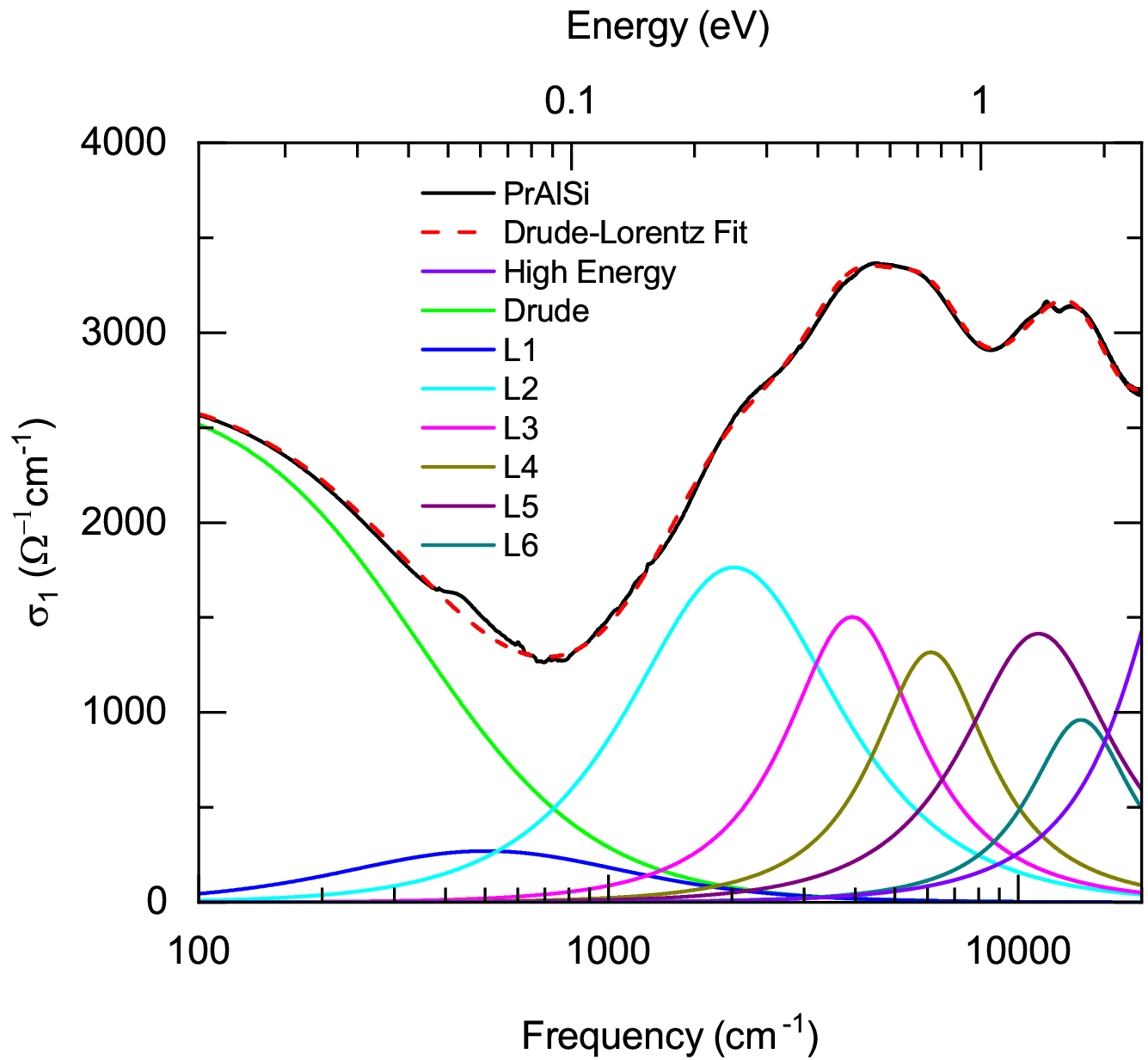}
	\caption{Drude-Lorentz fit of the optical conductivity $\sigma_1$ of PrAlSi at room temperature. All Drude and Lorentz contributions in the measured range are plotted separately, while the ``High-Energy'' contribution represents the sum of all higher-energy Lorentz peaks.}
	\label{fig.fittingcontributions}
\end{figure}

For quantitative analysis of the optical response functions, a Drude-Lorentz fitting model with one Drude component and six
Lorentz oscillators was applied for all five $R$AlSi compounds. As an example, we show in Fig.\ \ref{fig.fittingcontributions} the fit of the optical conductivity $\sigma_1$ for PrAlSi together with the various fitting contributions.
According to Hall resistivity measurements the main charge carrier type in the material LaAlSi are electrons~\cite{Su.2021}, which justifies the implementation of only one Drude oscillator term in the fitting model. For other $R$AlSi compounds, the contribution of both electron and hole pockets to the transport, i.e., the multiband nature, has been pointed out. Nevertheless, we keep the same Drude-Lorentz fitting model (one Drude term only) for all studied materials for consistency reason. We also point out that for none of the compounds it was necessary to include a second Drude term in the model, in order to obtain a very good fit of both reflectivity and optical conductivity data simultaneously.

\begin{figure}[t]
	\includegraphics[width=0.9\linewidth]{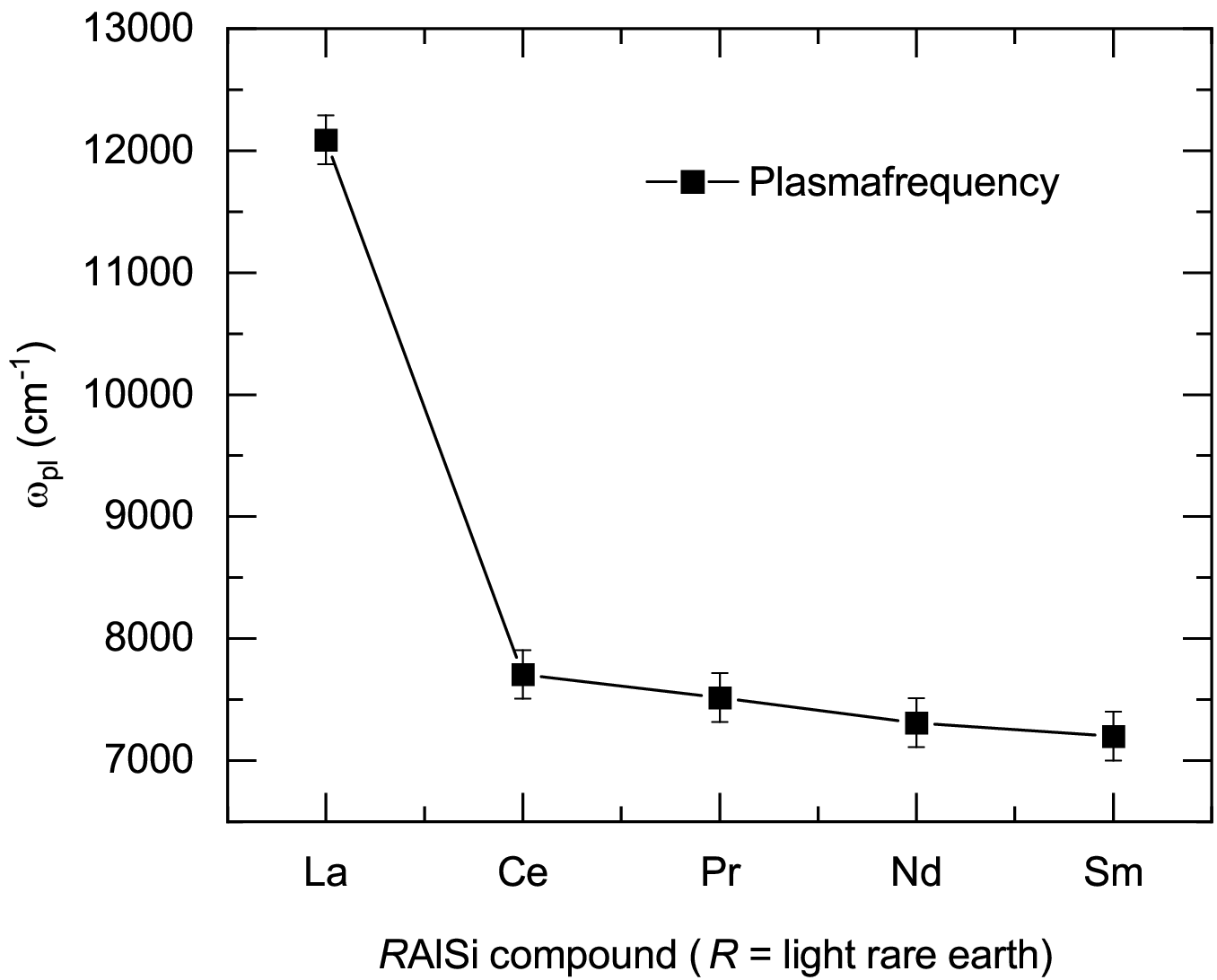}
	\caption{Plasma frequency $\omega_{pl}$ obtained from the spectral weight of the Drude term for all studied $R$AlSi compounds at room temperature.}
	\label{fig.plasmafrequency}
\end{figure}

The optical conductivity spectrum has a similar profile for all five compounds, however, there are quantitative differences: In LaAlSi the Lorentz oscillators L1 and L2 are shifted to lower energies as compared to the other compounds, implying slight differences in the electronic band structure close to $E_F$. The most pronounced differences are observed in the low-energy range, where the Drude term due to itinerant carrier excitations is located.
The plasma frequency $\omega_{\mathrm{pl}}$, as obtained from the spectral weight of the Drude term, is displayed in Fig.\ \ref{fig.plasmafrequency} as a function of rare earth element.
In comparison, the compounds CeAlSi, PrAlSi, NdAlSi, and SmAlSi have a very similar value of the plasma frequency ($\sim$7500\,\cm), LaAlSi exhibits a much higher value ($\sim$12000\,\cm).
%This behaviour follows the tendency as obtained from %$\omega_{\mathrm{pl}}^{\mathrm{scr}}$, where the %contributions of the material in its whole are taken into %account, rather than only the free charge carriers, which is %the case for the (unscreened) plasma frequency 5$\omega_{\mathrm{pl}}$.
%Thus, except for LaAlSi, all the other four compounds exhibit %a similar value in both cases.
It is important to note, that the metallic characteristics of the studied $R$AlSi compounds, as evidenced by the pronounced Drude term, is rather surprising, since electronic band structure calculations reveal an almost vanishing electronic density of states at the Fermi energy \cite{Piva.2023}.

We furthermore note that the values of the screened plasma frequency calculated from the Drude spectral weight
(taking into account the $\epsilon_1$ value at around 1500~cm$^{-1}$) are slightly higher than the values
as obtained from the zero crossing of $\epsilon_1$ (see inset of Fig.\ \ref{fig.epsilon1}). Such a discrepancy has been reported in the literature in several cases \cite{Li.2007,Li.2007a}, and is due to the choice of the frequency for extracting
the $\epsilon_1$ value and the error bar of the optical parameters like the Drude spectral weight.

\begin{figure}[t]
	\includegraphics[width=1\linewidth]{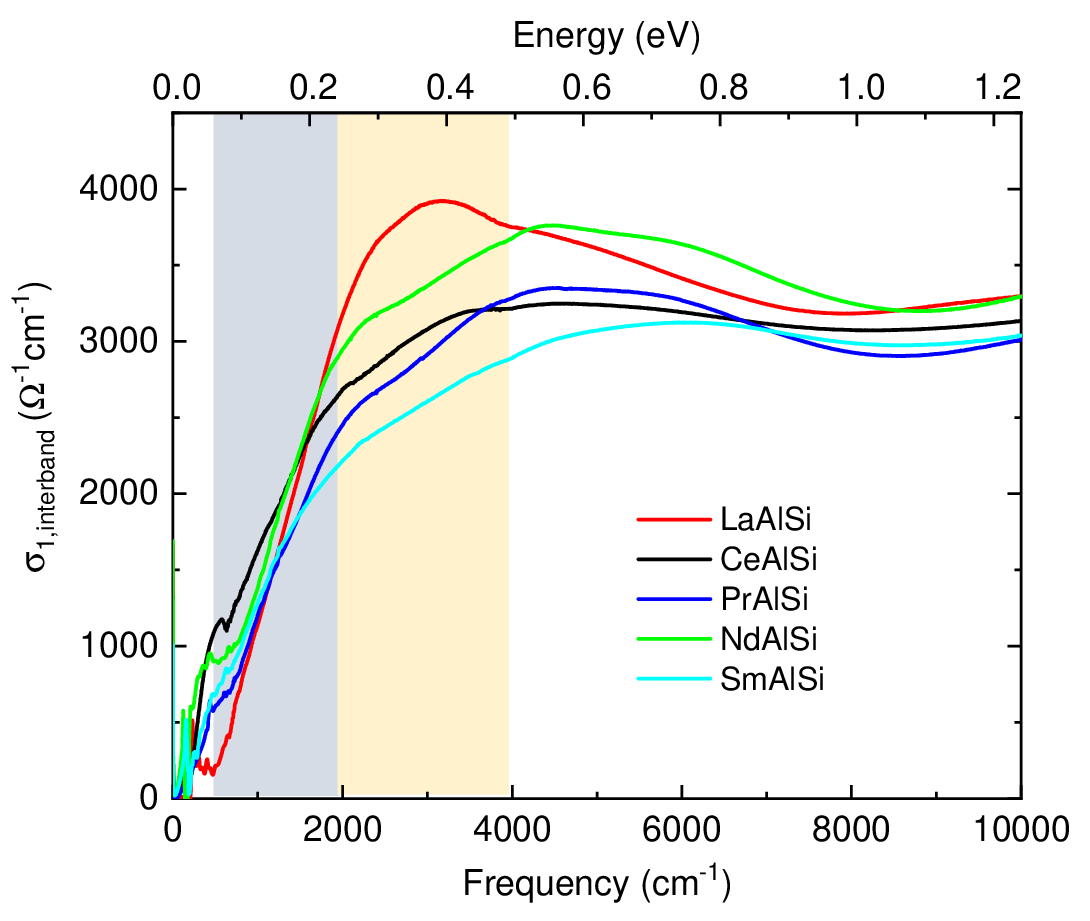}
	\caption{Interband optical conductivity of all studied $R$AlSi compounds at room temperature, as obtained by subtracting the Drude term from the total optical conductivity spectrum. The two regimes with the linear frequency dependences in $\sigma_{1,interband}$ are indicated by blue and orange areas, respectively. Please note that the marked areas only roughly highlight the two regimes, as there are quantitative differences among the studied materials regarding the limiting frequencies.}
	\label{fig.interband}
\end{figure}

For all studied compounds one notices that the optical conductivity $\sigma_1$ follows an $\omega$-linear behavior above
$\sim$1000~cm$^{-1}$, i.e., in the frequency range of the interband excitations. For further analysis and interpretation of
the interband transitions, we subtracted the Drude term from the total optical conductivity spectrum $\sigma_1$ and thus
obtained the interband conductivity $\sigma_{1,interband}$, which are plotted in Fig.\ \ref{fig.interband}.

\begin{figure}[t]
	\includegraphics[width=1\linewidth]{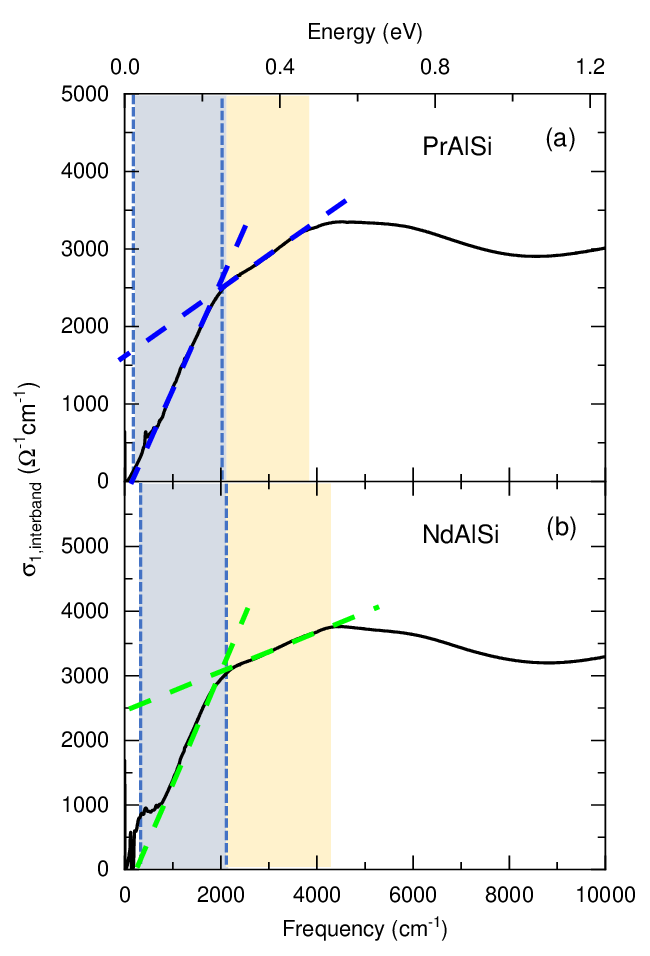}
	\caption{Interband optical conductivity spectra of (a) PrAlSi and (b) NdAlSi from 0 to 10000\,\cm at room temperature. In both spectra, two linear fits are inserted (dashed lines),  highlighting the characteristic features of type-II Weyl nodes.
The two regimes with the linear frequency dependences in $\sigma_{1,interband}$ are indicated by blue and orange areas, respectively, and limited by vertical, dashed lines.}
	\label{fig.linearfit}
\end{figure}

There are two linear regimes in $\sigma_{1,interband}$, as shown for PrAlSi and NdAlSi in Fig.\ \ref{fig.linearfit} as examples.
The steep linear increase in $\sigma_{1,interband}$ at low frequencies  ($\omega$$>$$\Omega_1$) is followed by another linear frequency dependence for $\omega$$>$$\Omega_2$, however, with a smaller slope, leading to a kink in the spectrum. This higher-energy linear frequency dependence (above $\Omega_2$) extrapolates to a finite $\sigma_{1,interband}$ value in the $\omega$ $\rightarrow$0 limit. The two regimes with the linear frequency dependences in $\sigma_{1,interband}$ are indicated in Fig.\ \ref{fig.linearfit} by blue and orange areas, respectively. For LaAlSi the higher-energy regime is least developed among the studied materials.

A linear-in-frequency behavior of the interband optical response is a characteristic fingerprint for the existence of Dirac and Weyl fermions in condensed matter.
Namely, the interband optical response of such linear Dirac/Weyl cones is predicted to follow a linear dependence in frequency $\omega$ for a three-dimensional (3D) system according to  \cite{Tabert.2016b,Ashby.2014,Hosur.2012,Timusk.2013}
\begin{equation}  \label{conductivity}
\sigma_1(\omega)=\frac{N e^2}{24 \pi \hbar v_F} \cdot \omega \cdot \Theta(\omega - 2 E_F) \quad ,
\end{equation}
where $N$ is the number of Weyl points (lifting the spin degeneracy) and v$_F$ the average Fermi velocity.
A possible shift of the Dirac node away from the Fermi energy is taken into account by the Heaviside step function $\Theta$, describing the Pauli blocking edge. For 3D Weyl semimetals, one or two quasilinear regions with different slopes are predicted \cite{Tabert.2016a,Tabert.2016b,Ashby.2014}, as described in more detail below.
The characteristic linear frequency dependence of the interband optical response was indeed observed in the 3D Dirac semimetal ZrTe$_5$ and in the 3D Weyl semimetal TaAs:
While for ZrTe$_5$ a linear behavior in $\sigma_1$ was observed in the low-frequency range \cite{Chen.2015}, for TaAs two $\omega$-linear components with different slopes were found \cite{Xu.2016}. In TaAs, the two $\omega$-linear components in the $\sigma_1$ spectrum are visible already at room temperature and persist down to low temperatures \cite{Kimura.2017}.

\begin{table*}
\caption{\label{Table-parameters} Frequencies $\Omega_1$ and $\Omega_2$, chemical potential $\mu$, tilting parameter $\hat{w}$ of the Weyl nodes, and average Fermi velocity $v_F$ of $R$AlSi compounds at room temperature, and for LaAlSi at room temperature and 5 K.}
\begin{ruledtabular}
\begin{tabular}{cccccc}
material       & $\Omega_1$ (cm$^{-1}$)  & $\Omega_2$ (cm$^{-1}$)  &  $\mu$ (cm$^{-1}$) & $\hat{w}$ & $v_F$ (10$^4$ ms$^{-1}$) \\
\hline
LaAlSi (295 K, 5 K)    & 412    &   2238        & 505 &1.45  & 36  \\
\hline
CeAlSi (295 K)        & 0 &   1957     & 0    & 1 & 61 \\
\hline
PrAlSi  (295 K)       & 148 &   1933    & 160    & 1.17 & 45 \\
\hline
NdAlSi  (295 K)       & 230 &   2072 & 259    & 1.45 & 46 \\
\hline
SmAlSi  (295 K)       & 69 &   1622 & 72    & 1.09 & 60 \\
\end{tabular}
\end{ruledtabular}
\end{table*}

According to theoretical predictions~\cite{Tabert.2016a},
for a time-reversal symmetry breaking WSM the Weyl cones are shifted along the direction in $k$-space, whereas for a space-inversion breaking WSM the cones are shifted in energy but stay situated at the same position in $k$-space. The latter situation results in an optical conductivity spectrum, where two linear-in-frequency ranges with different slopes are formed. The first linear slope is caused by excitations within the same Dirac cone, which requires less energy for electronic interband excitations in comparison. When the energy is high enough, these excitations can also be found between the two shifted Weyl cones. Thus, two linear ranges with different slopes are observed~\cite{Tabert.2016a}. Accordingly, from the number of linear-in-frequency regimes in $\sigma_1$ one could in principle distinguish between time-reversal symmetry breaking and space-inversion breaking WSMs.

On the other hand, also a tilting of the Weyl cones in reciprocal space can cause two linear-in-frequency regimes in the optical conductivity spectrum. The difference between type-I and type-II WSMs was described by Carbotte \cite{Carbotte.2016} based on the degree of tilting of the Weyl cones, given by the tilting parameter $\hat{w}$. Without tilting, the value $\hat{w}$=0 is assigned. For a small tilt, the values of $\hat{w}$ vary between $0$ and $1$. These states describe the case of a type-I WSM according to Ref.\ \cite{Carbotte.2016}. For a strong tilt, where the value of $\hat{w}$ exceeds $1$, a type-II WSM is formed. The value $\hat{w}$=1 describes the case for a tilt of $\pi/2$ into the plane perpendicular to the energy axis~\cite{Carbotte.2016}.
A type-II WSM is also referred to as an overtilted Weyl cone in energy-momentum space.
Based on the optical conductivity, the two types of WSMs can be differentiated by looking at the linear ranges in the spectrum: For a type-I WSM the linear-in-frequency behavior at frequencies $\omega$$>$$\Omega_2$ can be extrapolated to zero frequency, so that it crosses the origin. This is different for type-II WSM: These are also described by two linear slopes in the optical conductivity, where the extrapolation of the linear range at $\omega$$>$$\Omega_2$ yields an interception of the $y$-axis at a finite conductivity value. Furthermore, the slope of the higher-energy linear range ($\omega$$>$$\Omega_2$) is expected to decrease with an increasing tilt of the Weyl-nodes~\cite{Carbotte.2016}.
Please note that Ref.\ \cite{Carbotte.2016} considers Weyl nodes shifted away from the Fermi energy (see below).

According to the optical data obtained in this study, two linear ranges with different slopes are found in the $\sigma_{1,interband}$ spectra of the studied $R$AlSi materials (see Fig.\ \ref{fig.linearfit}). Furthermore, the $\omega$$\rightarrow$0 extrapolation of the linear slope in the higher-energy regime shows the crossing of the $y$-axis at a finite conductivity value, indicating that the investigated materials of the $R$AlSi compound family are in fact type-II WSMs formed by the space-inversion symmetry breaking. The data therefore suggest that the studied materials are type-II Weyl semimetals, with mainly overtilted $\hat{w} > 1$ Weyl nodes in the energy-momentum space (see also below). However, it is important to note, that in the case of LaAlSi both type-I and type-II Weyl nodes have been found and reported at $\sim$0.1~eV above $E_F$ and exactly at $E_F$, respectively~\cite{Su.2021}. Indeed, for LaAlSi the higher-energy linear-in-frequency regime in the $\sigma_1$ spectrum (see Fig.\ \ref{fig.reflectivity}) is less developed, which suggests a significant contribution of type-I Weyl nodes to excitations in the optical response. We also note that in the closely related compounds NdAlSi, CeAlSi, and LaAlGe 40 type-I and type-II Weyl nodes in the vicinity of the Fermi level have been theoretically predicted upon inclusion of spin-orbit coupling \cite{Wang.2022,Sakhya.2023,Xu.2017}.

In a recent study, optical conductivity data of the related materials LaAlGe and CeAlGe revealed the typical optical signatures of type-II Weyl fermions, namely two linear frequency dependences in the interband optical conductivity \cite{Corasaniti.2021}.
Interestingly, for PrAlGe only a single linear frequency dependence in $\sigma_{1,interband}$ was observed \cite{Yang.2022}, which is a typical signature for type-I Weyl semimetals, as discussed above.
For all three materials (LaAlGe, CeAlGe, PrAlGe), these signatures appear at much lower energies as compared to $R$AlSi, hence being not well separated from the free charge carrier excitations, and were much less developed at room temperature.
For CeAlGe and PrAlGe, the role of f-electrons has been pointed out, namely mediating electronic correlations (Kondo coupling between
f-electrons and conduction bands), which cause a renormalization of the topological bands, with the reduction of the Fermi velocity and the increase in the effective mass of charge carriers at low temperature \cite{Corasaniti.2021,Yang.2022}. Therefore, CeAlGe and PrAlGe have been suggested as correlated Weyl-Kondo systems. In contrast, LaAlGe does not possess f electrons and therefore is expected to be non-correlated \cite{Corasaniti.2021}.

Comparing the optical response of $R$AlSi and $R$AlGe materials, we note that $all$ $R$AlSi compounds of the present study show a similar profile of the optical conductivity spectrum at room temperature, i.e., above the magnetic ordering, revealing the typical signatures of type-II Weyl fermion excitations. In particular, we cannot confirm a type-I Weyl fermion behavior for PrAlSi in contrast to PrAlGe. Like LaAlGe, LaAlSi is expected to be non-correlated as there are no f electrons, consistent with the observed highest Drude spectral weight (squared unscreened plasma frequency) among the $R$AlSi compounds (see Fig.\ \ref{fig.plasmafrequency}).
Following the analysis of the interband optical conductivity spectra in Ref.\ \cite{Corasaniti.2021}, we can extract the tilting parameter $\hat{w}$ of the Weyl cones, the chemical potential $\mu$, and the average Fermi velocity $v_F$. According to Ref.\ \cite{Carbotte.2016}, the parameters $\mu$ and $v_F$ can be obtained from the limiting frequencies $\Omega_1$ and $\Omega_2$ using the equations $\Omega_1$=2$\mu$/(1+$\hat{w}$) and $\Omega_2$=2$\mu$/($\hat{w}$-1). The so-obtained parameters for the studied $R$AlSi compounds at room temperature are listed in Table \ref{Table-parameters}.
For all studied materials $R$AlSi the tilting parameter $\hat{w}$ is 1 or greater than 1, suggesting an overtilt of the Weyl cones.
Furthermore, from the slope of the linear-in-frequency behavior of the interband optical conductivity at $\omega$$>$$\Omega_2$ (see Fig.\ \ref{fig.linearfit}) the Fermi velocity can be estimated according to \cite{Carbotte.2016}
\begin{equation}  \label{Weylconductivity}
\sigma_1(\omega)=\frac{N e^2}{24 \pi \hbar v_F} \cdot \frac{1+3\hat{w}^2}{4\hat{w}^3}  \cdot \omega \quad .
\end{equation}
For all studied compounds we set N=40 \cite{Wang.2022,Sakhya.2023,Xu.2017}.
The so-obtained values of the Fermi velocity $v_F$ are included in Table \ref{Table-parameters}. The values of $v_F$ for all compounds lie in the range (35 - 61) $\cdot$$10^4$ ms$^{-1}$, in reasonable agreement with results from Shubnikov-de Haas oscillation measurements on SmAlSi and LaAlSi at low temperature \cite{Xu.2022,Cao.2022,Su.2021}. For LaAlSi without f electrons and hence no low-temperature magnetic ordering, we also extracted the
low-temperature parameters from corresponding optical data and found them to be independent of temperature.
Finally, it is important to note that the $v_F$ values of $R$AlSi are a factor of $\sim$10 higher than for LaAlGe and CeAlGe \cite{Corasaniti.2021}, suggesting that the $R$AlSi materials are less correlated than $R$AlGe.

\section{Conclusion}
In conclusion, we studied the optical response of the noncentrosymmetric compounds $R$AlSi ($R$=La, Ce, Pr, Nd, Sm) at room temperature, in order to investigate their electronic structure and electronic excitations in the paramagnetic phase, in particular regarding the dependence on the rare-earth element $R$. Over the whole studied frequency range, the reflectivity and optical conductivity spectra show only minor differences between the materials CeAlSi, PrAlSi, NdAlSi, and SmAlSi, which all exhibit magnetic order at low temperatures.
The main differences can be found in the frequency range below 7500\,\cm, where the free charge carrier excitations and low-frequency interband transitions occur.
Especially the nonmagnetic LaAlSi compound stands out, as it exhibits the strongest metallic character according to its (screened) plasma frequency.
Thus, mostly the electronic band structure close to the Fermi level seems to be affected by the variation of the rare-earth element $R$. Furthermore, two quasi-linear ranges with different slopes are observed in the low-frequency optical conductivity spectrum, which indicates the existence of Weyl nodes in the vicinity of E$_F$. From the analysis of the linear-in-frequency behavior of the optical conductivity we can conclude that the Weyl cones are mainly of type-II and are overtilted.

\begin{acknowledgments}
This work was supported by the Deutsche Forschungsgemeinschaft (DFG, German Research Foundation) -- TRR 360 -- 492547816.
YQ would like to acknowledge the support by the National Natural Science Foundation of China (Grant Nos.\ 52272265, U1932217, 11974246, 12004252) and the National Key R \& D Program of China (Grant No\ 2018YFA0704300).

\end{acknowledgments}

\end{document}